\documentclass[aps,prl,reprint,groupedaddress]{revtex4-2}

\usepackage{bm}
\usepackage{graphicx}
\usepackage{mathtools}
\usepackage{amsmath}
\usepackage{physics}
\usepackage{siunitx}
\usepackage{xcolor}
\usepackage{float}

\begin{document}
	
	\title{Augmenting the sensing performance of entangled photon pairs through asymmetry}
	
	\author{Yoad Michael}
	\affiliation{Bar-Ilan University, 5290002 Ramat-Gan, Israel}
	\author{Isaac Jonas}
	\affiliation{Bar-Ilan University, 5290002 Ramat-Gan, Israel}
	\author{Leon Bello}
	\affiliation{Bar-Ilan University, 5290002 Ramat-Gan, Israel}
	\author{Mallachi-Ellia Meller}
	\affiliation{Bar-Ilan University, 5290002 Ramat-Gan, Israel}
	\author{Eliahu Cohen}
	\affiliation{Bar-Ilan University, 5290002 Ramat-Gan, Israel}
	\author{Michael Rosenbluh}
	\affiliation{Bar-Ilan University, 5290002 Ramat-Gan, Israel}
	\author{Avi Pe'er}
	\affiliation{Bar-Ilan University, 5290002 Ramat-Gan, Israel}
	\email{avi.peer@biu.ac.il}
	
	\date{\today}
	
	\begin{abstract}
		We analyze theoretically and experimentally cases of asymmetric detection, stimulation and loss within a quantum nonlinear interferometer of entangled pairs. We show that the visibility of the SU(1,1) interference directly discerns between loss on the measured mode (signal), as opposed to the conjugated mode (idler). This asymmetry also affects the phase sensitivity of the interferometer, where coherent seeding is shown to mitigate losses that are suffered by the conjugated mode, therefore increasing the maximum threshold of loss that permits sub shot-noise phase detection. Our findings can improve the performance of setups that rely on direct detection of entangled pairs, such as quantum interferometry and imaging with undetected photons.
	\end{abstract}
	
	
	\maketitle
	The main goal in quantum interferometry is to measure small phase shifts with very high sensitivity, beyond what could normally be achieved with classical light schemes that are limited by shot-noise \cite{Caves1981, Ou2012}. The major resource for sub-shot-noise interferometry is squeezed light \cite{Bondurant1984, Caves1985, Schnabel2017, Taguchi2020, Bishop1988, Lawrie2019, Schoenbeck2017}, which is simple to generate by parametric down conversion and is widely utilized, e.g. by LIGO and VIRGO \cite{Grote2013, Aasi2013, McCuller2020, Miller2015, Barsotti2018, Acernese2019, Acernese2020}. \textcolor{black}{A common way of harnessing squeezed light is injection into the unused port of a classical Mach-Zehnder interferometer (MZI) \cite{Pezze2008}, yielding a phase sensitivity that is enhanced by the degree of squeezing \cite{Wodkiewicz1985}.} 
	
	An alternative and very useful interferometric approach is to replace the two beam-splitters of the MZI with optical parametric amplifiers (OPAs) \cite{Cerullo2003}. An OPA produces entangled photon pairs (or squeezed light), each comprised of two correlated spectral modes known as the signal and idler, that have a well-defined phase relation with respect to the pumping field. The OPA can be thought of as a two-mode squeezer that amplifies one quadrature of the combined signal-idler field, and attenuates the other quadrature \cite{Bello2020, Schumaker1985}. 
	
	Placing two consecutive OPAs with a phase shift in-between yields a phase-dependent output (See Fig. 1a), similarly to the interference output of a beam-splitter, with one key advantage over its classical counterpart: if no loss exists between the two OPAs, the optical state is not mixed with external vacuum, yielding sub shot-noise phase sensitivity with no need for external squeezing \cite{Caves2020, Guo2018}. This configuration is known as the SU(1,1) interferometer \cite{Yurke1986, Chekhova2016, Ou2020}.
	
		\begin{figure} \label{figure1}
		\centering
		\includegraphics[width=7.5cm]{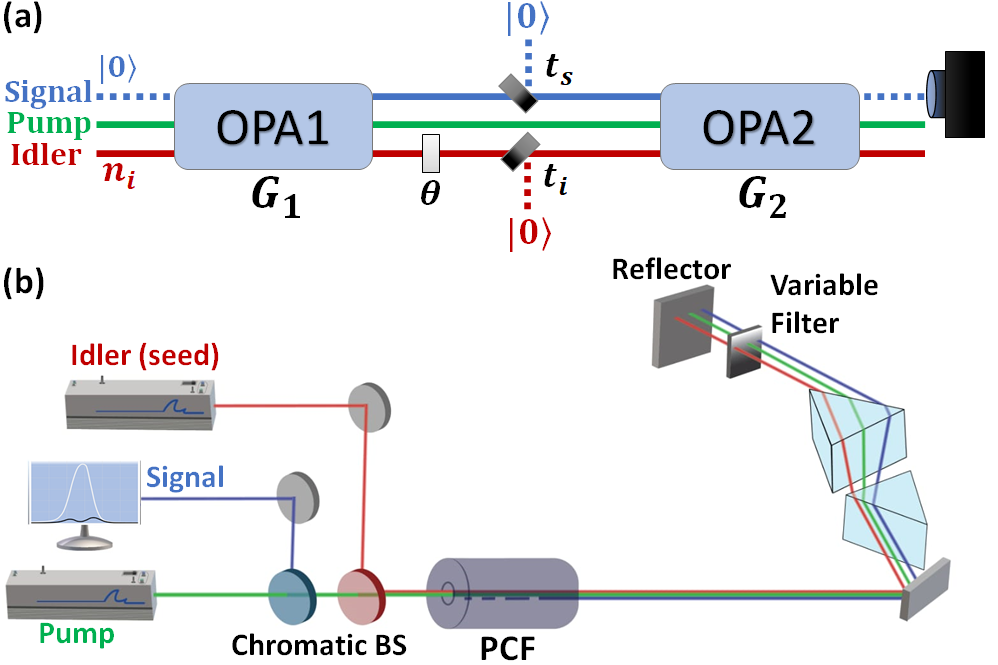}
		\caption{Conceptual and experimental overview of the system. (a) Nonlinear SU(1,1) interferometer with losses.  A relative phase shift $\theta$ is applied between the OPAs, the modes experience independent photon losses, and the nonlinear interference occurs at OPA2. (b) Experimental realization using a photonic crystal fiber (PCF) as both OPAs: First pass (left to right) generates entangled photon pairs, while the nonlinear interference occurs on the second pass (right to left). The signal is seperated from the pump and idler using a chromatic beam-splitter (BS) and measured by a spectrometer.}
	\end{figure}

	In addition to the improved sensitivity, the SU(1,1) interferometer is a highly flexible scheme that was previously utilized in various forms: A balanced configuration where the intensity is measured directly \cite{Anderson2017} or through parity detection \cite{Li2016}, a truncated configuration \cite{Gupta2018, Prajapati2019} where the second OPA is replaced by standard homodyne detection, a non-balanced configuration where the second OPA is set to a higher gain for broadband parametric homodyne measurement \cite{Shaked2018, Li2019} and overcoming detection loss \cite{Manceau2017, Manceau2018, Li2017}, etc. The improved signal-to-noise due to the squeezing was also harnessed for various other applications, such as Raman spectroscopy \cite{Michael2019, Prajapati2021} and atomic force microscopy \cite{Pooser2020}. The SU(1,1) interferometer can operate in multiple regimes: quantum or classical \cite{Vered2015}, spontaneous or stimulated \cite{Plick2010, Li2014}, with optical loss degrading the squeezing \cite{Marino2012, Hu2018, Xin2016}.
	
	While the SU(1,1) interferometer is typically discussed in the context of squeezing-enhanced detection, a different aspect of the interference was utilized for quantum imaging with undetected photons \cite{Lemos2014}: The IR spectrum of a molecular sample that absorbs light at the idler wavelengths, was inferred from the visibility of nonlinear interference measured at the signal wavelength \cite{Kalashnikov2016, Paterova2020, Kviatkovsky2020}, which is far more optically-accessible. A similar setup was also used for optical coherence tomography \cite{Vanselow2020}. In these schemes, the symmetry between the two modes that comprise the pair is inherently broken: while the idler experiences photon loss, it is the signal that is measured. This effect of asymmetry between the modes is critical both for quantum imaging and squeezing-enhanced applications, yet is usually not elaborated upon and is given as experimental framework.     
	
	Here we explore theoretically and experimentally the effect of asymmetric optical loss, measurement and seeding on the SU(1,1) interference of entangled photon pairs. We show that this interference is critically affected by whether the mode (signal or idler) that experiences loss is also the one that is measured, or the conjugate one. We show that the visibility can be twice as high when losses are applied on the signal (the measured mode) compared to the idler (conjugated). The phase-sensing performance of the interferometer is also affected: In the spontaneous regime, higher phase sensitivity is observed when losses are applied on the measured mode than on the conjugated mode. When the conjugated mode is coherently seeded, it withstands higher degree of loss while maintaining sub shot-noise phase sensitivity, \textcolor{black}{where we show a 1.5dB improvement over the unseeded case. We demonstrate a tailored visibility-transmission curve for optimized visibility at high signal losses.} 
	
	We start with the mathematical description of the OPA. An OPA is a nonlinear gain medium that generates entangled signal and idler pairs from a \textcolor{black}{strong and narrowband} pump \cite{MacLean2018, Yang2020, Peextquotesingleer2005}. The quantum state of the entangled pairs is described by the input-output relations of the field operators before and after passing the OPA $\hat{a}_{s,i}^{(out)} = \cosh(G)\hat{a}_{s,i}^{(in)}+\sinh(G){\hat{a}^\dagger_{i,s}}{}^{(in)}$, where $\hat{a}_{s,i}^{(in,out)}$ are the field operators of the signal/idler at the input and output of the OPA, and $G$ is the parametric gain. \textcolor{black}{While the operators of the signal and idler each represent a single spectral mode, this description is also valid for OPAs that produce broadband non-degenerate pairs, as long as adjacent pairs are not spectrally entangled~\cite{Chen2014}, for example due to a broadband pump.} 
	
	\begin{figure} \label{figure2}
		\centering
		\includegraphics[width=6.5cm]{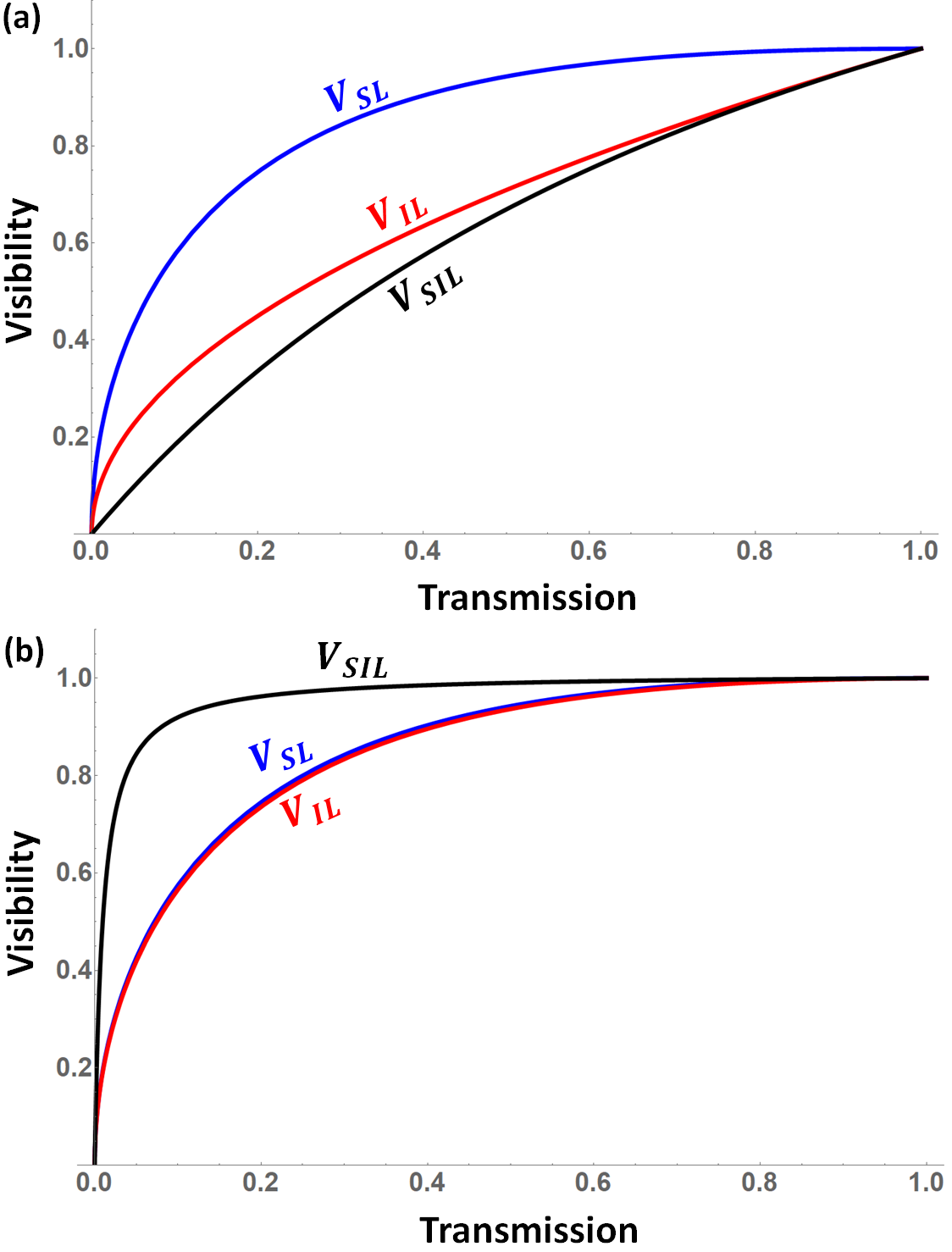}
		\caption{The theoretical visibility of nonlinear interference as a function of the transmission ($t_{s,i}^2$) of the signal ($V_{SL}$, blue), idler ($V_{IL}$, red) and equal transmission ($V_{SIL}$, black), for (a) the spontaneous $n_i=0$ and (b) idler-stimulated $n_i=50$ cases, at the low gain regime of $G_1=G_2=0.1$. The interference is measured at the signal mode. For the spontaneous case, the highest visibility is observed when losses are applied on the measured mode ($V_{SL}$), highlighting the asymmetry to loss between the measured and conjugated modes. For the stimulated case, the loss asymmetry between the two modes breaks ($V_{SL}=V_{IL}$), while equal loss ($V_{SIL}$) achieves the highest visibility, in correspondence to Eqs.~3-5.}
	\end{figure}
	
	We place two sequential OPAs with gains $G_{1,2}$. A phase shift $\theta$ and optical loss \textcolor{black}{(transmission coefficients $t_{s,i}$) are applied between OPAs} (See Fig. 1a). The most general expression for the photon number at the signal output of the interferometer is (see the Supplemental Material for derivation):     
	\textcolor{black}{\begin{equation} \label{eq1}
		\begin{aligned}
			\left<\hat{N}_s\right> &= (n_i+1)\left(\beta \cos (\theta ) t_i t_s +  {\lambda}_{2,1} t_i^2 + {\lambda}_{1,2} t_s^2 \right)
			\\ &+\delta_2 \left(1-t_i^2\right),
		\end{aligned}
	\end{equation}}
	\textcolor{black}{where we used a shorthand notation for the parametric gains: $\beta \equiv \frac{1}{2} \sinh \left(2 G_1\right) \sinh \left(2 G_2\right)$,  ${\lambda}_{2,1} \equiv \sinh ^2\left(G_2\right) \cosh ^2\left(G_1\right)$, ${\lambda}_{1,2} \equiv \sinh ^2\left(G_1\right) \cosh ^2\left(G_2\right)$,  $\delta_{2} \equiv \sinh ^2\left(G_2\right)$, and seeded the interferometer with a coherent state of $n_i$ average photons at the idler port (and vacuum for the signal).} To understand the physical meaning of the different terms in this general expression, let us first consider a simplified case with no losses $t_s=t_i=1$, and phase of $\theta=0,\pi$ (corresponding to constructive and destructive interference). The expression then reduces to \textcolor{black}{${\left<\hat{N}_s\right>}_{0, \pi}=(n_i+1) \sinh ^2\left(G_1\pm G_2\right)$}, with the seed photons $n_i$ acting as a classical stimulation for the pair generation.  When losses are introduced, the interference terms (first three terms of Eq.1) are transmitted, but a non-interfering term of \textcolor{black}{$\left(1-t_i^2\right) \delta_{2}$} is introduced from the spontaneous amplification of the quantum vacuum by OPA2. Note that this term arises from loss on the conjugated mode - hinting at the asymmetry to loss.
	
	To understand the consequences of this term, we proceed to define the visibility (contrast) of interference as $V=\frac{\left<\hat{N}_s\right>_{\theta=0}-\left<\hat{N}_s\right>_{\theta=\pi}}{\left<\hat{N}_s\right>_{\theta=0}+\left<\hat{N}_s\right>_{\theta=\pi}}$, where $V=1$ indicates full distinction between constructive and destructive interference.
	
	\textcolor{black}{We start with the most general expression for the visibility of the SU(1,1) interference, which can be derived from Eq.~1 following the definition of the visibility:}
	
	\textcolor{black}{\begin{equation} \label{eq2}
			V = \frac{\beta \left(n_i+1\right) t_i t_s}{{\lambda}_{1,2}\left(n_i+1 \right) t_s^2 + \delta_2 \left(1+t_i^2 \left( \left(n_i + 1 \right) \delta_1 - 1 \right) \right)},
	\end{equation}}
	
	\textcolor{black}{where $\delta_1 \equiv \cosh^2 (G_1)$.}
		
	We move on to three simplified cases of interest: The visibility due to losses only on the signal ($V_{SL}$), only on the idler ($V_{IL}$), and equal loss on both ({$V_{SIL}$}), which can all be derived from Eq. 2. We start with balanced amplifiers $G_1=G_2=G$. When loss is applied only on the signal ($t_i=1$) the visibility is 
	\begin{equation} \label{eq3}
		\begin{split}
			V_{SL} = \frac{2 t_s}{t_s^2+1},
		\end{split}
	\end{equation} 
	which is independent of the gain and stimulation, and approaches $100\%$ for $t_s \rightarrow 1$, as shown by the blue curve in Figs.~2a and 2b. Next, for losses only on the idler ($t_s=1$), the visibility is
	\begin{equation} \label{eq4}
		\begin{split}
			V_{IL} = \frac{2 (n_i+1) t_i \cosh ^2(G)}{(n_i+1) \left(t_i^2+1\right) \cosh ^2(G)+1-t_i^2}.
		\end{split}
	\end{equation} 
	Generally $V_{IL}<V_{SL}$ (shown by the red curve in Fig. 2) indicating that higher visibility is observed when applying loss on the measured mode (signal) than on the conjugated mode (idler), with $V_{IL} \rightarrow \frac{2 t_i}{t_i^2+1}$ for high values of $n_i$ (Fig. 2b). Intuitively, since the signal and idler undergo nonlinear interference together, one might expect $V_{SL}=V_{IL}$, yet the question of which of the modes is measured (in our case, the signal) breaks the symmetry between the entangled pairs and discerns $V_{SL}$ from $V_{IL}$ \textcolor{black}{in Fig.~2a, up to a factor of two}. 
	
	The third interesting case is equal (symmetric) loss on both the signal and idler ($t_s=t_i=t$):
	\begin{equation} \label{eq5}
		\begin{split}
			V_{SIL} = \frac{2 (n_i+1) t^2 \cosh ^2(G)}{2(n_i+1) t^2 \cosh ^2(G)+1-t^2},
		\end{split}
	\end{equation} 
	Which for the spontaneous case has the lowest visibility (black curve in Fig. 2a), yet approaches $100\%$ for large values of $n_i$, even for a low transmission (Fig. 2b). 
	
	\textcolor{black}{Many important schemes in quantum imaging, illumination and tomography employ the basic structure of the SU(1,1) interferometer, indicating that better understanding of the asymmetric effects has important implications to all of them. We can deduce that in terms of optimizing the visibility, an optical implementation of the unseeded SU(1,1) should minimize the losses on the measured mode, while for the seeded interferometer the losses should be kept symmetrical.}
	
	While high visibility is critical for the mentioned applications, it is not a direct indicator of optimal phase sensitivity, which is a key feature of the SU(1,1) interferometer. We now consider how the asymmetry affects the phase sensitivity. Using error propagation analysis, the sensitivity (minimum detectable phase shift) is defined as $\Delta \theta^2=\frac{\left<\hat{N}_s^2\right>-\left<\hat{N}_s\right>^2}{|\frac{d}{d\theta}|_{\theta=\theta_0}\left<\hat{N}_s\right>|^2},$
	where $\theta_0$ is the phase-working point, around which small phase shifts $\Delta \theta$ are measured.
	 
	Generally, the phase sensitivity of SU(1,1) depends on the gain of each OPA, the internal photon loss and the phase-working point. For the ideal case (no loss, balanced amplifiers, coherently seeded idler) the sensitivity is:
	\begin{equation} \label{eq6}
		\Delta \theta^2=\frac{1}{(1+n_i)\sinh^2(2G)}=\frac{1}{4(1+n_i)(N_{sq}^2+N_{sq})},
	\end{equation}
	where $N_{sq}=\sinh^2(G)$ reflects the parametric gain in terms of the number of spontaneously generated photon-pairs. While the seeding term $n_i$ has classical scaling, it is multiplied by the sub shot-noise term of $N_{sq}^2$. The coherent seed reinforces the low flux of squeezed photons with a high flux of coherent photons that stimulate the generation of pairs. This stimulation does not increase the squeezing, but it ``upscales'' the quantum effect, greatly increasing its applicability in real-life sensing applications that inherently require a high flux of photons. 
	\begin{figure} \label{figure3}
		\centering
		\includegraphics[width=6.5cm]{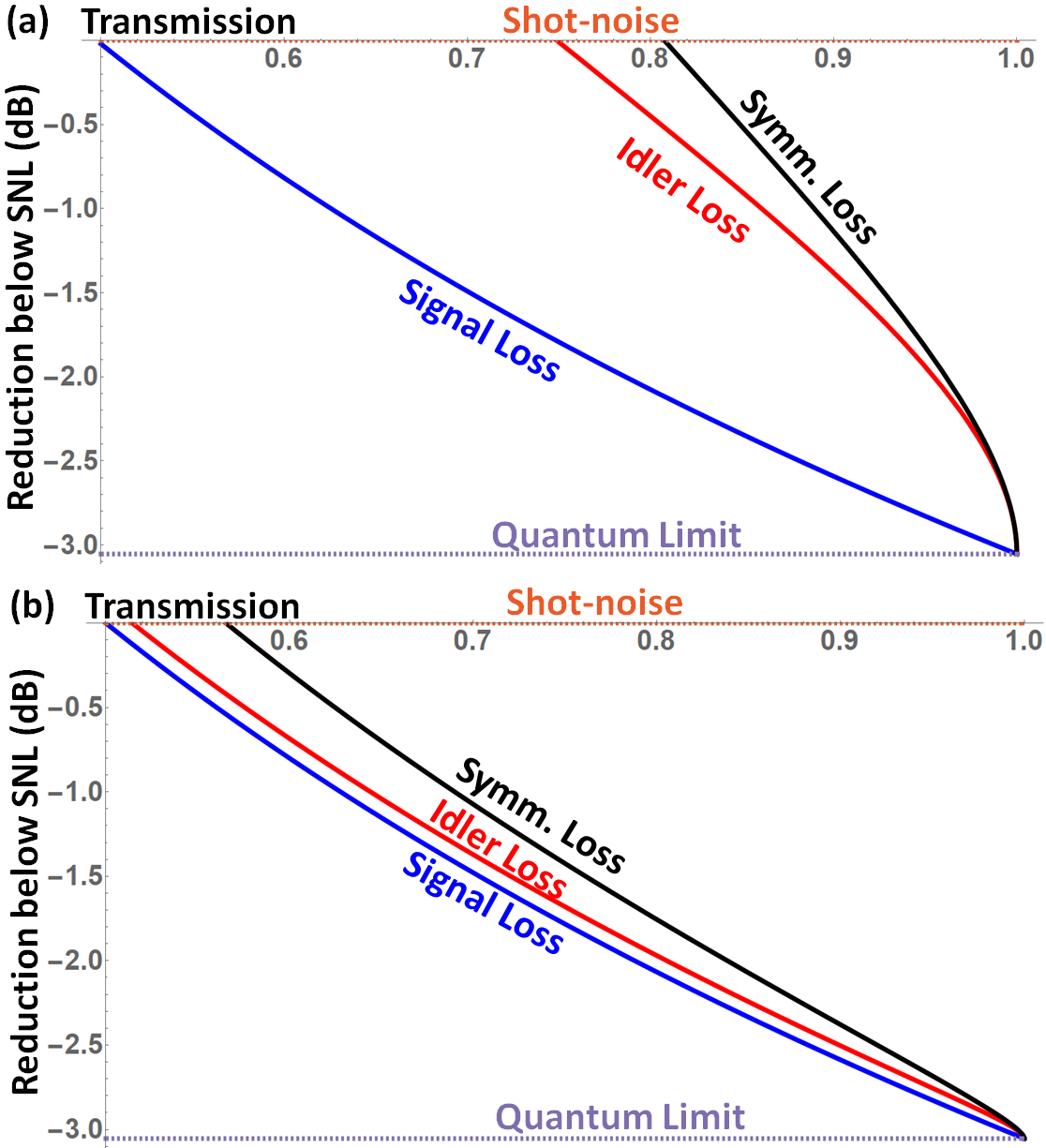}
		\caption{The theoretical phase sensitivity (relative to the shot-noise) as a function of the transmission ($t_{s,i}^2$) of the signal (blue), idler (red) and equal transmission (black), for (a) the spontaneous $n_i=0$ and (b) stimulated $n_i=50$ cases, with $G_1=G_2=0.1$. The dashed-orange curve represents the classical shot-noise level (values below this level indicate squeezing), and the dashed-gray curve represents the limit of an ideal SU(1,1) interferometer. For the spontaneous case (a) loss on the measured mode (signal) achieves higher phase sensitivity than loss on the conjugated mode (idler). For the stimulated case (b) the gap between the curves closes, indicating that coherent stimulation on the conjugated mode increases its phase sensitivity in the presence of loss.}
	\end{figure} 
	
	Following the error-propagation method, we calculate the optimal phase sensitivity in the presence of loss: asymmetric (signal or idler) and symmetric (both). We normalize the phase-sensitivity relative to the shot-noise level \textcolor{black}{($1/N_s$, where $N_s$ is the number of signal photons inside the interferometer)}, and explore both the spontaneous (Fig.~3a) and coherently seeded (Fig.~3b) regimes. 
	
	For the spontaneous SU(1,1) interferometer, a clear difference between the three cases is observed: signal-only loss (blue) shows the highest robustness to loss compared to idler-only (red) and symmetric (black) losses, which is the worst case. When a coherent seed is added, the signal-only case is unchanged (and remains optimal), but the gap between the curves closes. \textcolor{black}{In our example, at $75\%$ idler transmission, the phase sensitivity was shot-noise limited for the spontaneous case, while the stimulated case showed 1.5dB improvement over the shot-noise. This indicates that for phase sensing, the stimulation can compensate for losses on the conjugated mode.}
	
	\textcolor{black}{Our results are directly applicable to any practical application of the SU(1,1) interferometer for sub shot-noise sensing, including phase shifts, squeezing-enhanced spectroscopy and microscopy, which are all operating in the vicinity of moderate optical loss. In that regard, it is better to measure the mode that experiences the adverse losses. If the conjugated mode does experience loss, it can be compensated by seeding the interferometer. This has implications for quantum illumination schemes \cite{Tan2008, Shapiro2020, Nair2020}, where one mode is used to sense a weakly reflecting target (experiencing major loss) with the other mode retained as reference.}
		
	Looking back at Fig.~2 and Fig.~3, we can see that the visibility and phase sensitivity seem to coincide in their response to asymmetric losses (blue and red lines), with signal loss being optimal, and idler loss improving by the stimulation. In contrast, the case of symmetric (equal) loss has the highest visibility when stimulated, yet it achieves the worst sensitivity. This strengthens our previous statement that high visibility is not a direct indicator of optimal phase sensitivity and squeezing. 
	
	\textcolor{black}{The relevant parameters should therefore be chosen based on the specific application - For example, quantum imaging requires high visibility and therefore benefits from a seeded configuration with symmetrical loss, while sub shot-noise quantum sensing should minimize the losses on the measured mode regardless of the seed.}
	
	\textcolor{black}{We now discuss our experimental setup, in which we tailored the visibility-transmission curve to distinguish between losses on the signal (measured) and idler (conjugated). We expand our analysis to include an imbalanced interferometer ($G_1 = 0.45, G_2=0.2$) and different starting conditions for losses on the signal (initial transmission of $52 \%$) and idler (initial transmission of $42 \%$).} \textcolor{black}{In order to understand the theoretical behavior under these conditions, it is best to refer to Eq.~(2), which takes into account mixed losses and gains. The unbalanced configuration also causes a distinguishing effect on the measured and conjugated modes, as we show in our experiment and elaborate in the Supplemental Material.}  
		
	We constructed a four-wave mixing based SU(1,1) interferometer, as shown in Fig. 1b. Our pump was a Ti:Sapphire laser at 786nm (100mWatt average power, 8ps pulses), which was coupled into a photonic-crystal fiber (PCF) as the OPA to generate signal and idler pairs. We separated the signal, idler and pump beams in space using a prism pair, which also served as a phase control (using its variable dispersion). We applied optical loss independently on the signal and idler using two variable transmission filters. We then reflected the light back into the PCF for a second pass, where the nonlinear interference between signal-idler pairs and pump occurred.
	
	\begin{figure} \label{figure4}
		\centering
		\includegraphics[width=6.5cm]{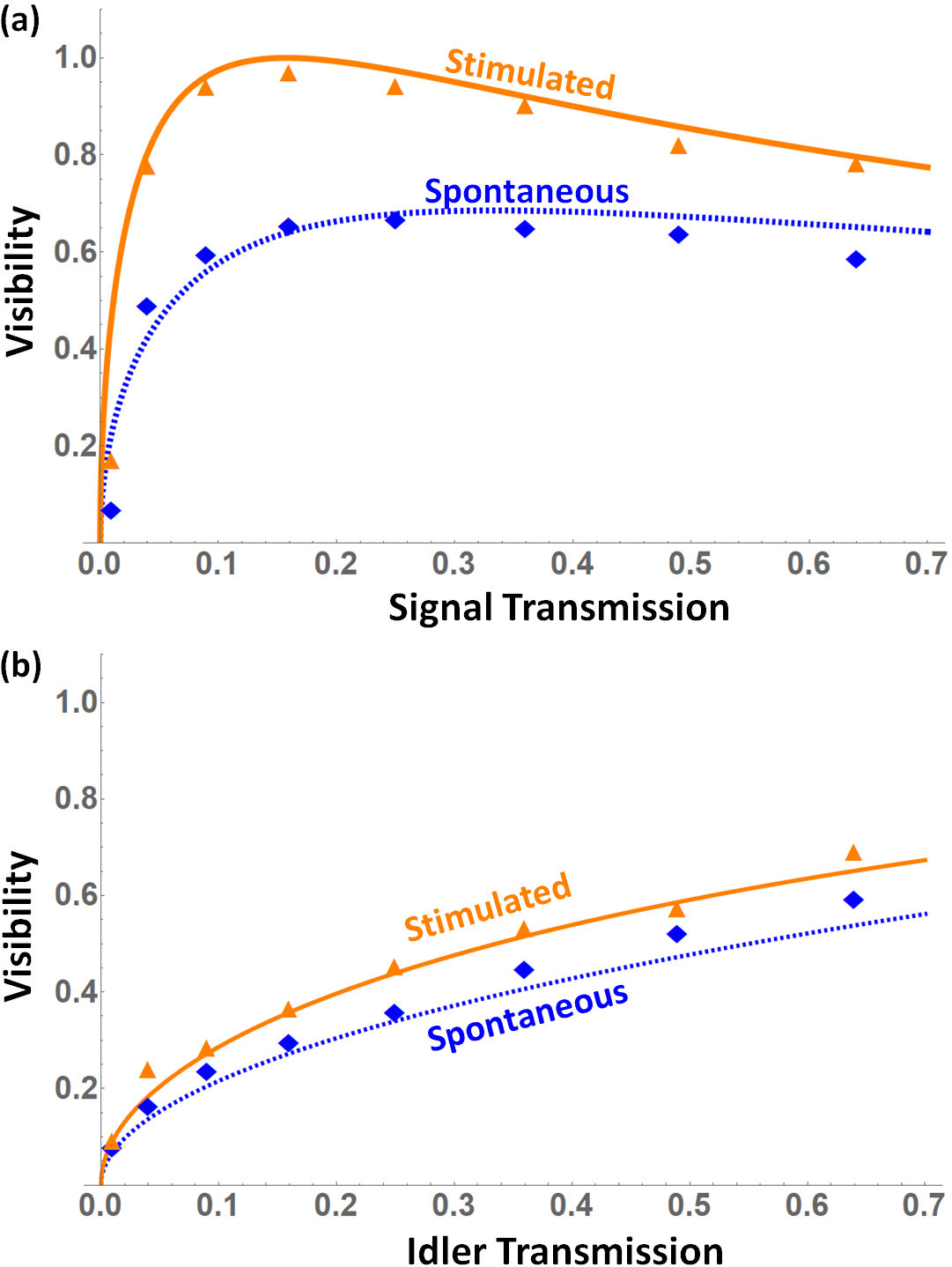}
		\caption{Experimental and theoretical visibility as function of the transmission of the signal (a) and idler (b), for the spontaneous ($n_i=0$, blue, dashed) and  stimulated cases ($n_i=10^4$, orange, solid), with $G_1=0.45, G_2=0.2$. \textcolor{black}{The initial transmission of the signal and idler is $52\%$ and $42\%$, on top of which we further reduce the transmission of the signal (a) and idler (b).} Each point in the graph is calculated by two consecutive measurements of constructive and destructive interference, with the triangle/diamond markers representing the experimental data. The experimental error is up to $ \pm 0.02$.}
	\end{figure} 
	
	The spectrum of the signal was then measured using a spectrometer, and the visibility was extracted from two consecutive measurements of constructive and destructive interference. We measured the visibility as a function of the transmission of the signal (Fig. 4a) and the idler (Fig. 4b) separately for both the spontaneous ($n_i=0$, blue) and the strongly stimulated ($n_i=10^4$, orange) cases. \textcolor{black}{Our choice of a strong, coherent seed at 878nm generated a signal (at 712nm) that is well above the noise floor of our detector, even with a short integration time of $10 \mu s$, which is beneficial for real-time applications of quantum imaging and sensing \cite{Basset2021}.}
	
	Our experimental results are shown in Fig. 4, where the inherent asymmetry in the losses can be observed from the visibility. In Fig. 4a, where the optical transmission of the signal is varied, the visibility increases for the stimulated case as the transmission of the signal decreases, reaching $92\%$ at a signal transmission of $16\%$. \textcolor{black}{The increased visibility at that point is attributed to two different effects: The first is loss-balancing between the signal and idler, and the second is the gain difference between OPA1 and OPA2, which pushes the optimal visibility further towards lower signal (measured mode) transmission.} On the other hand, increasing the loss of the idler (conjugated), which had higher internal loss in the first place, always degrades the visibility (Fig. 4b) for both the spontaneous and stimulated cases. 
	
	\textcolor{black}{A tailored visibility-transmission curve is of particular interest in applications that try to distinguish between losses that are specific to either the signal or the idler, for example absorption spectroscopy. If we take the stimulated case of Fig. 4a, we can see that if a weakly absorbing sample is placed within the interferometer, it would cause an increase in the visibility (if the sample absorbs the signal) or a decrease (absorbs the idler). It is also relevant for quantum imaging and tomography with undetected photons, in which it would be possible to maximize the visibility by tuning the asymmetric parameters (measurement, loss, seed intensity and gain ratio). The visibility-transmission curve can just as easily be tailored for increased visibility at low idler transmission, as we show in the Supplemental Material.}
	
	\textcolor{black}{An interesting freedom that we did not explore here is the bandwidth of the pump, which could create a multi-mode entanglement \cite{Fabre2020} by coupling between neighboring pairs. This effect is unsuitable for applications that measure fine spectral structures, but it possibly allows broadband phase sensing beyond the shot-noise limit.} 
	
	To conclude, we presented the analysis of asymmetry between the entangled photons in the SU(1,1) interferometer. We have shown that the inherent loss asymmetry between the measured and conjugated mode can be used to optimize the visibility and phase sensitivity. We experimentally demonstrated a tailored visibility-transmission curve that clearly distinguishes between loss on the measured and conjugated modes. We also highlight the robustness of the SU(1,1) interferometer, which was already shown to be effective against detection losses in past research \cite{Manceau2017, Frascella2021} and here expanded to the case of asymmetric internal losses. 
	
	\textit{Acknowledgments.} We thank Yuri Kaganovskii and Joseph Kantorovitsch for producing the variable filters and providing major insight. E.C. and A.P. were supported by the Israel Innovation authority under grant 7002 and 73795. E.C. acknowledges support from the Quantum Science and Technology Program of the Israeli Council of Higher Education, from FQXi and from the Pazy foundation.

\end{document}